\documentclass[twocolumn,showpacs,prd]{revtex4}
\usepackage{graphicx}
\usepackage{dcolumn}
\usepackage{bm}
\begin{document}

\title{Space-Like Motions of Quantum Zero Mass Neutrinos}
\author{Allan Widom and John Swain}
\affiliation{Physics Department, Northeastern University, Boston MA USA}
\author{Yogendra Srivastava}
\affiliation{Physics Department \& INFN, University of Perugia, Perugia IT}

\begin{abstract}
Recent experimental reports of super-luminal velocity neutrinos moving between 
Geneva and Gran Sasso in no way contradict the special relativity considerations 
of conventional quantum field theory. A neutrino exchanged between Geneva and Gran 
Sasso is both virtual and space-like. The Lorentz invariant space-like distance 
$L$ and the Lorentz invariant space-like four momentum transfered $\varpi $ between 
Geneva and Gran Sasso can be extracted from experimental data as will be shown in this work.  
\end{abstract}

\pacs{14.60.Lm, 13.15.+g, 03.65.Pm}

\maketitle

\section{Introduction \label{intro}}

There have been recent large group experimental 
reports\cite{MINOS:2007,OPERA:2011} that neutrinos have been 
observed to travel at super-luminal speeds, 
i.e. at speeds faster than light in an approximately inertial 
frame of reference. The neutrinos were reported to travel at super-luminal speeds  
from CERN in Geneva to CNGS in Gran Sasso. Should the experimental assertions  
of super-luminal neutrino motion turn out to have sufficient verification, 
then important and deep theoretical understanding of relativistic physics 
will eventually accrue. Our purpose is not to probe the detailed experimental 
procedures leading to reports of faster than light speed neutrinos. In what follows, 
we merely discuss the theoretical implications of such ultra-relativistic processes.     

Let us first note within the context of quantum field theory, the fact that a neutrino  
moves in a super-luminal manner by no means implies a breakdown of the Einstein 
special relativity or of Lorentz invariance. It merely means that the experimental 
Lorentz invariant quantities must be chosen with care. The neutrino moving between 
Geneva to Gran Sasso is virtual as shown in the Feynman diagram of FIG.\ref{fig1} in 
Sec.\ref{rvp}. (i) The Lorentz invariant space-like displacement of the virtual neutrino
is denoted by $L$. From the data $L\sim 6\ {\rm kilometer}$. (ii) The Lorentz invariant 
space-like momentum transfer carried by the virtual neutrino between Geneva and Gran Sasso 
is denoted by $\varpi $. From the data $\varpi\sim 100\ {\rm MeV/c}$. 

In Sec.\ref{slv} the special relativity kinematics of super-luminal velocities are discussed. 
The method of extracting the Lorentz invariant space-like displacement $L$ from experimental 
data is provided. The particle reactions in Geneva and in Gran Sasso are explored. 
In Sec.\ref{rvp} the virtual neutrino between Geneva and Gran Sasso are visualized. The 
virtual neutrino exchanges a Lorentz invariant space-like virtual neutrino momentum transfer 
$p^2=\varpi^2>0$ [We use the metric $\eta^{oo} = -1$, i.e. (+1,+1,+1,-1)]. 
The method of extracting the Lorentz invariant space-like $\varpi $ from 
experimental data is provided. In Sec.\ref{nwf} the nature of the quantum wave function of 
a virtual neutrino is explored.

\section{Super-Luminal Kinematics\label{slv}}

For a particle moving with velocity  
\begin{equation}
{\bf v}=\frac{d{\bf r}}{dt},
\label{slv1}
\end{equation}
a sub-luminal velocity defines the Lorentz invariant proper time of motion as 
\begin{eqnarray}
c^2d\tau ^2=c^2 dt^2-|d{\bf r}|^2 \ \ {\rm for}
\ \ |{\bf v}| < c,
\nonumber \\  
d\tau=dt\sqrt{1-\left( \frac{|{\bf v}|}{c} \right)^2}.
\label{slv2}
\end{eqnarray}
For a particle moving at super-luminal velocities, 
the Lorentz invariant proper space of motion is 
\begin{eqnarray}
ds^2=|d{\bf r}|^2 -c^2 dt^2 \ \ {\rm for}
\ \ |{\bf v}| > c,
\nonumber \\ 
ds=d|{\bf r}|\sqrt{1-\left( \frac{c}{|{\bf v}|} \right)^2}.
\label{slv3}
\end{eqnarray}

\subsection{Space-Like Displacement \label{sld}}

For a super-luminal velocity which is uniform in space and time, 
\begin{equation}
v=\frac{|{\bf r}_1 - {\bf r}_2 |}{|t_1 - t_2 |} > c,
\ \ \ {\rm with}\ \ \ r=|{\bf r}_1 - {\bf r}_2 |
\label{slv4}
\end{equation}
the Lorentz invariant space-like displacement is as in Eq.(\ref{slv3}); i.e. 
\begin{eqnarray}
L^2=| {\bf r}_2 -{\bf r}_1 |^2-c^2|t_2-t_1|^2,
\nonumber \\ 
L=r\sqrt{1-\left(\frac{c}{v}\right)^2}=
r\sqrt{\left(\frac{v+c}{v}\right)\left(\frac{v-c}{v}\right)}.
\label{slv5}
\end{eqnarray}
While an experimental measurement of {\it r} and {\it v} 
depend on that Lorentz frame of reference to be named the 
laboratory frame, the space-like displacement {\it L} is independent 
of which Lorentz frame of reference is chosen for a physics analysis. 

What is a bit more subtle is the analysis of the direction of time. It 
is well known that Feynman\cite{Feynman:1949} analyzed the notion of a particle as going 
forward in time and an anti-particle as going backwards in time. If the particle 
goes space-like with a displacement {\it L}, as in Eq.(\ref{slv4}), then in 
one Lorentz frame the particle may go $1 \to 2$ and in another 
Lorentz frame the anti-particle may go $2 \to 1$\cite{Weinberg1972}. For example, 
if an experimental group reports that a super-luminal neutrino went from 
Geneva to Gran Sasso in some laboratory Lorentz frame of reference, then 
in another Lorentz frame of reference a super-luminal anti-neutrino went from 
Gran Sasso to Geneva.  There is simply no Lorentz invariant way to time order 
space-like separated events. 

All of these considerations do not require that Einstein be abandoned. 
One can in a relativistic invariant way simply quote the Lorentz invariant 
space-like displacement {\it L}. The reported super-luminal European 
neutrino or anti-neutrino transport asserts a space-like displacement of  
\begin{math} L\sim 6\ {\rm kilometer} \end{math} with 
values depending on laboratory distance scales 
\begin{math} r \sim 730\ {\rm kilometer} \end{math} and laboratory 
neutrino energy. 

It is crucially important to discuss these matters in terms 
of physical quantities which are Lorentz invariants. Sub-luminal particles 
are described in terms of Lorentz invariant quantities such as mass {\it m} 
determined by \begin{math}  p^2=-m^2c^2 \end{math}. One should not 
forget that energy {\it E} is merely one component of a four momentum 
\begin{math} p_\mu =({\bf p},-E/c)  \end{math}. When particles or anti-particles 
are super-luminal, the four momentum definition perhaps becomes somewhat 
subtle but the requirement of a Lorentz invariant description remains sacrosanct.
  
\subsection{Particle Reactions \label{pr}}

Consider the decay of an anti-pion into an anti-muon plus a muon 
neutrino, 
\begin{equation}
\pi^+ \to \mu^+ +\nu_\mu . 
\label{slv6}
\end{equation}
The neutrino is detected by a nuclear weak interaction via 
\begin{equation}
\nu_\mu +N_i \to N_f + \mu^-
\label{slv7}
\end{equation}
Altogether, the reaction amounts to the creation of a muon pair changing the 
internal state of the nucleus {\it N}; i.e. 
\begin{equation}
\pi^+ +N_i \to N_f+\mu^+ +\mu^-.  
\label{slv8}
\end{equation}
There is no neutrino in Eq.(\ref{slv8})  because the neutrino is merely virtual. 
In a profound Appendix B in a paper\cite{Feynman:1950} wherein Feynman gave a derivation from 
quantum field theory of the diagram rules, he asserts that what looks real on a 
short time scale appears virtual on a long time scale. Let us apply this discussion 
to the muon pair production Eq.(\ref{slv8}) with the 
\begin{math} \mu^+  \end{math} in Geneva and the 
\begin{math} \mu^-  \end{math} in Gran Sasso. It appears clear that a 
neutrino or anti-neutrino had to make the cross country trip whether 
or not one considers such neutrino or anti-neutrino motions as real or virtual. 
 
\section{Real and Virtual Processes \label{rvp}}

Let us first consider the neutrino or anti-neutrino four momentum {\it p}. 
Regarded as a real pion decay process in Eq.(\ref{slv6}), the four momentum 
\begin{equation}
p=p(\pi^+)-p(\mu^+)
\label{rvp1}
\end{equation} 
Regarded as a real neutrino absorption in Eq.(\ref{slv7}), the four momentum 
\begin{equation}
p=p(N_f)+p(\mu^-)-p(N_i).
\label{rvp2}
\end{equation} 
Equating the above two expressions for the momentum {\it p} yields the  
conservation of four momentum pair production Eq.(\ref{slv8}) for which
\begin{equation}
p(\pi^+)+p(N_i)=p(N_f)+p(\mu^+)+p(\mu^-).
\label{rvp3}
\end{equation} 
While it is nice that four momentum in Geneva + Gran Sasso is all conserved, 
that fact alone does not answer questions concerning a neutrino that no 
longer appears in the momentum balance. 

\subsection{Classical Neutrino Four Momenta \label{cnfm}}

For a classical {\em real zero mass neutrino} the four momentum obeys 
\begin{equation}
p^2=|{\bf p}|^2-(E/c)^2=0\ \ \ \ \Rightarrow 
\ \ \ E=c|{\bf p}|.
\label{cnfm1}
\end{equation}
The velocity based on the classical Hamilton equation 
\begin{equation}
{\bf v}=\frac{\partial E}{\partial {\bf p}}=\frac{c{\bf p}}{|{\bf p}|}=c{\bf n}
\ \ \ \Rightarrow \ \ \ v=|{\bf v}|=c.
\label{cnfm2}  
\end{equation}
Thus, a classical zero mass particle moves at light speed. This no longer 
holds true in quantum mechanics wherein a zero mass particle can move 
off the light cone.

\subsection{Quantum Neutrino Propagation \label{qnp}}

\begin{figure}[tp]
\scalebox {0.70}{\includegraphics{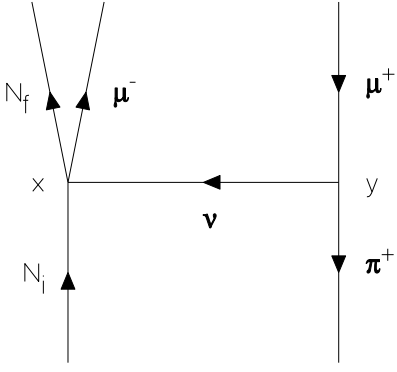}}
\caption{Shown is the lowest order neutrino exchange diagram between an event 
in Geneva at $y$ and an event in Gran Sasso at $x$ according to Eqs.(\ref{slv6}), 
(\ref{slv7}) and (\ref{slv8}). The virtual neutrino exchange may be space-like; i.e. 
$(x-y)^2>0$.}
\label{fig1}
\end{figure}

The Schwinger action\cite{Schwinger:1989} corresponding to a massless neutrino 
source spinor  
\begin{math} \eta \end{math} reads 
\begin{equation}
W=\hbar \int \int \bar{\eta}(x)S(x-y)\eta(y)d^4x d^4y,
\label{qnp1}
\end{equation}
wherein the zero mass spin one-half particle propagator 
\begin{math}  S(x-y) \end{math} obeys 
\begin{equation}
-i\gamma^\mu \partial_\mu S(x-y)=\delta (x-y).
\label{qnp2}
\end{equation} 
The one neutrino exchange amplitude for the particle interaction 
Eq.(\ref{slv8}) is then 
\begin{eqnarray}
{\cal A}(\pi^+ +N_i \to N_f+\mu^+ +\mu^-)=i \int d^4x \int d^4y 
\nonumber \\ 
\left< N_f \mu^- \right|\bar{\eta}(x)\left|N_i \right>
S(x-y) \left< \mu^+ \right|\eta(y)\left|\pi^+ \right>,
\label{qnp3}
\end{eqnarray}
corresponding to the Feynman diagram in FIG.\ref{fig1} .
Solving Eq.(\ref{qnp2}) in the form 
\begin{equation}
S(x-y)=i\gamma^\nu \partial_\nu D(x-y),
\label{qnp4}
\end{equation}
yields the propagator \begin{math} D(x-y)  \end{math} obeying
\begin{equation}
-\partial^\mu \partial_\mu D(x-y)=\delta(x-y).
\label{qnp5}
\end{equation}
Upon Fourier transformation and with the Feynman virtual particle 
boundary conditions 
\begin{eqnarray}
D(x-y)=\int \tilde{D}(Q)e^{iQ\cdot (x-y)}\frac{d^4Q}{(2\pi )^4}\ ,
\nonumber \\ 
\tilde{D}(Q)=\frac{1}{Q^2-i0^+}\ ,
\nonumber \\ 
D(x-y)=\frac{i}{4\pi^2}
\left(\frac{1}{(x-y)^2 +i0^+}\right). 
\label{qnp6}
\end{eqnarray}
With the notation for principal part left implicit, one may write 
\begin{eqnarray}
\tilde{D}(Q)=\frac{1}{Q^2}+i\pi \delta(Q^2)  ,
\nonumber \\ 
D(x-y)=\frac{1}{4\pi^2}\left[\pi \delta((x-y)^2)
+i \left(\frac{1}{(x-y)^2}\right)\right] . 
\label{qnp7}
\end{eqnarray}
The neutrino four momentum in the diagram of FIG.\ref{fig1} 
is \begin{math} p=\hbar Q  \end{math}. 

Note that the part of the propagator wherein the neutrino 
is off the energy shell, i.e. \begin{math} p^2\ne 0 \end{math} 
and \begin{math} E\ne \pm c|{\bf p}| \end{math}, 
gives rise to motions on the light cone 
i.e. \begin{math} (x-y)^2= 0 \end{math} 
and \begin{math} ct= \pm c|{\bf r}| \end{math}. 
In detail, with \begin{math} x-y=({\bf r},ct)  \end{math},  
\begin{eqnarray}
\int  \left[\frac{e^{iQ\cdot (x-y)}}{Q^2}   \right]
\frac{d^4Q}{(2\pi )^4}=\frac{\delta((x-y)^2)}{4\pi}=
\nonumber \\ 
\frac{\delta(r^2-c^2t^2)}{4\pi }=
\frac{1}{8\pi r}\left[ \delta(r-ct) +\delta (r+ct) \right] . 
\label{qnp8}
\end{eqnarray}
Similarly, the part of the propagator wherein the neutrino 
is on the energy shell, i.e. \begin{math} p^2 = 0 \end{math} 
and \begin{math} E = \pm c|{\bf p}| \end{math}, 
gives rise to motions off the light cone 
i.e. \begin{math} (x-y)^2 \ne 0 \end{math} 
and \begin{math} ct \ne \pm c|{\bf r}| \end{math}. In detail 
\begin{eqnarray}
\int  \delta(Q^2) e^{iQ\cdot (x-y)}  
\frac{d^4Q}{(2\pi )^4}=\frac{1}{4\pi^3 (x-y)^2}=
\nonumber \\ 
\frac{1}{4\pi^3 (r^2-c^2t^2)}=\frac{1}{4\pi^3 }\left(\frac{1}{L^2}\right).
\label{qnp9}
\end{eqnarray}
The uncertainty principle  
\begin{math} \Delta E\Delta t > \hbar/2 \end{math}
{\em forbids} the neutrino to be {\em both} on the 
energy shell \begin{math} p^2=0 \end{math} and on the light cone 
\begin{math} (x-y)^2=0 \end{math}. 

From Eqs.(\ref{qnp8}) and (\ref{qnp9}), it follows that neutrinos can move 
on the forward and backward light cones. It can move on time-like and space-like 
intervals. The quantum theory goes considerably beyond the classical causal retarded 
in time viewpoint employing the forward light cone  
\begin{equation}
D_{\rm ret}({\bf r},t)=\frac{1}{4\pi r}\delta (r-ct).
\label{qnp10}
\end{equation}
The quantum propagator can be computed from the retarded propagator by an analytic 
continuation in the imaginary time domain 
\begin{eqnarray}
\Delta({\bf r},t^2)=\frac{i}{\pi }\int_0^\infty 
\frac{t^\prime D_{\rm ret}({\bf r},t^\prime )dt^\prime }{{t^\prime }^2-t^2}
\ \ {\rm if}\ \ {\Im }m(t^2)<0, 
\nonumber \\ 
D(x-y)=\Delta({\bf r},t^2-i0^+) 
\ \ {\rm for}\ \ (x-y)=({\bf r},ct).
\label{qnp11}
\end{eqnarray}
The above method for computing acausal quantum propagation from classical causal 
propagation has wide validity\cite{srivastava:1987}.  

\subsection{Virtual Neutrino Four Momentum \label{vnfm}}

For the neutrino exchange diagram in FIG.\ref{fig1} when employed in four 
momentum space, the virtual neutrino can have a space-like four momentum, 
\begin{equation}
p^2=|{\bf p}|^2-(E/c)^2=\varpi^2>0,
\label{vnfm1}
\end{equation} 
so that the energy of a virtual neutrino $+$ or virtual anti-neutrino $-$ 
has the form 
\begin{equation}
E=\pm c\sqrt{|{\bf p}|^2-\varpi^2}\ .
\label{vnfm2}
\end{equation} 
The velocity of the virtual neutrino
\begin{equation}
{\bf v}=\frac{\partial E}{\partial {\bf p}}
\ \ \Rightarrow \ \ {\bf p}=\frac{E{\bf v}}{c^2}
\ \ \Rightarrow \ \ |{\bf v}|>c.
\label{vnfm3}
\end{equation} 
Finally,
\begin{eqnarray}
L\ {\rm and}\ \varpi \ {\rm are\ Lorentz\ invariant},
\nonumber \\ 
|{\bf r}|\ {\rm and}\ |{\bf p}| \ {\rm are\ Lorentz\ frame\ dependent},
\nonumber \\ 
\frac{L}{\varpi }=\frac{|{\bf r}|}{|{\bf p}|}\ .
\label{vnfm4}
\end{eqnarray}
Typically\cite{MINOS:2007,OPERA:2011} the reported values of the Lorentz invariants 
should read \begin{math} L\sim 6\ {\rm kilometer} \end{math} and  
\begin{math} \varpi \sim 100\ {\rm MeV/c} \end{math} in accordance with 
Eqs.(\ref{slv5}) and (\ref{vnfm4}).

\section{The Neutrino Wave Function \label{nwf}}

In terms of the Feynman scattering amplitude we may discuss the 
neutrino wave function \begin{math} \psi (x)  \end{math} .
The neutrino wave function at space-time point {\it x} due to a source event 
at space-time point {\it y} is given by the amplitude in FIG.\ref{fig1} or equivalently 
in Eq.(\ref{qnp3}) as 
\begin{equation} 
\psi(x)=\int S(x-y) \left< \mu^+ \right|\eta(y)\left|\pi^+ \right> d^4y.
\label{nwf1}
\end{equation}
Employing Eq.(\ref{qnp4}), the effective source 
\begin{equation} 
s(y)=i\gamma^\mu \partial_\mu \left< \mu^+ \right|\eta(y)\left|\pi^+ \right> ,
\label{nwf2}
\end{equation}
and integrating Eq.(\ref{nwf1}) by parts yields 
\begin{equation} 
\psi(x)=\int D(x-y) s(y) d^4y.
\label{nwf3}
\end{equation}
Note that the neutrino wave function obeys the massless Dirac equation only away 
from the source; i.e. invoking Eq.(\ref{nwf2}),
\begin{eqnarray} 
-i\gamma^\mu \partial_\mu \psi (x)=\left< \mu^+ \right|\eta(x)\left|\pi^+ \right>,
\nonumber \\ 
(i\gamma^\nu \partial_\nu)(-i\gamma^\mu \partial_\mu) \psi (x)=
-\partial^\mu \partial_\mu \psi(x)=s(x), 
\label{nwf4}
\end{eqnarray}
again leading to Eq.(\ref{nwf3}).

\section{Conclusion \label{conc}}

The reported super-luminal velocity neutrinos moving between 
Geneva and Gran Sasso do not imply a breakdown of the special relativity 
considerations of conventional quantum field theory if the neutrino motion 
between Geneva and Gran Sasso is regarded as being virtual. The two Lorentz 
invariant quantities describing the virtual neutrino are as follows:
(i) The space-like interval $L$ traveled by the virual neutrino is given by 
\begin{equation}
L=|{\bf r}|\sqrt{1-\left(\frac{c}{|{\bf v}|}\right)^2}. 
\label{conc1}
\end{equation}
(ii) Even for a massless neutrino, the four momentum of the neutrino can be 
space-like. Thereby the Lorentz invariant four momentum square obeys    
\begin{equation}
p^2=\varpi^2>0. 
\label{conc2}
\end{equation}
For the experimental reports at hand, the virtual neutrino travels 
$L\sim 6\ {\rm kilometer}$ with a momentum transfer of 
$\varpi \sim 100\ {\rm MeV/c} $.

\end{document}